# A Great Space Weather Event in February 1730

Hisashi Hayakawa (1, 2), Yusuke Ebihara (3, 4), José M. Vaquero (5), Kentaro Hattori (6), Víctor M. S. Carrasco (7), María de la Cruz Gallego (7), Satoshi Hayakawa (8), Yoshikazu Watanabe (9), Kiyomi Iwahashi (10), Harufumi Tamazawa (11), Akito D. Kawamura (11), and Hiroaki Isobe (4, 12)

1    Rutherford Appleton Laboratory, Chilton, Didcot, Oxon OX11 0QX, UK

*e-mail: hisashi.hayakawa@stfc.ac.uk

2    Graduate School of Letters, Osaka University, Toyonaka, 5600043, Japan (JSPS Research Fellow)

**e-mail: hayakawa@kwasan.kyoto-u.ac.jp

3    Research Institute for Sustainable Humanosphere, Kyoto University, Uji, 6100011, Japan

4    Unit of Synergetic Studies for Space, Kyoto University, Kyoto, 6068306, Japan

5    Departamento de Física, Universidad de Extremadura, E-06800 Mérida, Spain

6    Graduate School of Science, Kyoto University, Kyoto, 6068501, Japan

7    Departamento de Física, Universidad de Extremadura, E-06071 Badajoz, Spain

8    Faculty of Engineering, The University of Tokyo, 1130033, Tokyo, Japan

9    Oriental Astronomical Association, 6500021, Kobe, Japan

10    National Institute of Japanese Literature, Tachikawa, 1900014, Japan

11    Kwasan Observatory, Kyoto University, Kyoto, 6078471, Japan

12    Graduate School of Advanced Integrated Studies in Human Survivability, Kyoto University, Kyoto, 6068306, Japan

**ABSTRACT**

**Aims**. Historical records provide evidence of extreme magnetic storms with equatorward auroral extensions before the epoch of systematic magnetic observations. One significant magnetic storm occurred on February 15, 1730. We scale this magnetic storm with auroral extension and contextualise it based on contemporary solar activity.

**Methods**. We examined historical records in East Asia and computed the magnetic latitude (MLAT)

of observational sites to scale magnetic storms. We also compared them with auroral records in Southern Europe. We examined contemporary sunspot observations to reconstruct detailed solar activity between 1729 and 1731.

**Results**. We show 29 auroral records in East Asian historical documents and 37 sunspot observations.

**Conclusions**. These records show that the auroral displays were visible at least down to 25.8° MLAT throughout East Asia. In comparison with contemporary European records, we show that the boundary of the auroral display closest to the equator surpassed 45.1° MLAT and possibly came down to 31.5° MLAT in its maximum phase, with considerable brightness. Contemporary sunspot records show an active phase in the first half of 1730 during the declining phase of the solar cycle. This magnetic storm was at least as intense as the magnetic storm in 1989, but less intense than the Carrington event.

## 1. Introduction

It is well established that large sunspots frequently cause large solar eruptions, resulting in great magnetic storms with auroral displays, even in low-latitude areas (e.g. Daglis 2003; Willis et al. 2006; Odenwald 2015; Takahashi & Shibata 2017; Riley et al. 2018). The Carrington event in 1859 is one of the earliest flares connected to the greatest magnetic storm in the history of ground-based telescopic observations (Kimball 1960; Tsurutani et al. 2003; Cliver & Svalgaard 2004). On the very date of this event, Carrington (1859, 1863) recorded a great sunspot that was estimated to be up to 3000 millionths of a solar hemisphere (Cliver & Keer 2012; Hayakawa et al. 2016b). This great sunspot caused a white-light flare to bring great magnetic storms to the Earth, with low-latitude auroras up in worldwide sites up to 22-23° magnetic latitude (MLAT). Auroral displays were visible in sites such as Hawaii, the Caribbean Coast, and Southern Japan (Carrington 1859; Hodgson 1859; Neidig & Cliver 1983a,b; Kimball 1960; Tsurutani et al. 2003; Cliver & Svalgaard 2004; Farrona et al. 2011; Cliver & Dietrich 2013; Hayakawa et al. 2016b; Lakhina & Tsurutani 2016; Riley et al. 2018). The magnetic storm brought intense magnetic disturbances at low latitudes, with maximum negative intensity up to 1600 nT, recorded at Colaba (Tsurutani et al. 2003; Nevanlinna 2004, 2006; Ribeiro et al. 2011; Cliver & Dietrich 2013; Viljanen et al. 2014; Kumar et al. 2015). These magnetic storms severely affected the telegraph network in Europe and North America (Boteler 2006; Cliver & Dietrich 2013).

It is intriguing how frequently such extreme storms occur and how intense they can be, as such extreme events can be even more hazardous to a modern civilisation that relies upon electric infrastructure (Daglis 2003; Hapgood 2011, 2012; Odenwald 2015; Riley et al. 2018). The U. S. Research Council warns us that another Carrington storm in modern times will be catastrophic enough to cause disasters costing as much as 2 trillion USD (e.g. Baker et al. 2008). Fortunately, the recent extreme solar eruption in 2012 was a near miss, while this CME was considered to be as intense as the Carrington flare, and modern civilisation managed to avoid a potentially great catastrophe (Baker et al. 2013; Liu et al. 2014; Riley et al. 2018). While studies on great flares and the resultant great magnetic storms after the mid-19th century have been conducted recently (e.g. Allen et al. 1989; Silverman & Cliver 2001; Cliver & Svalgaard 2004; Shiokawa et al. 2002, 2005; Vaquero et al. 2008; Silverman 1995, 2006, 2008; Cliver & Dietrich 2013; Araki 2014; Cid et al. 2014; Viljanen et al. 2014; Kilpua et al. 2015; Vennerstrom et al. 2016; Lefèvre et al. 2016; Knipp et al. 2016; Hayakawa et al. 2016b; Lockwood et al. 2016; Love 2017), solar flares and the resultant magnetic storms before the Carrington event have not been well studied due to lack of systematic observations for solar flares and magnetic disturbances.

Recent works suggest that even more energetic flares can occur in our Sun over a longer time span. Studies in stellar physics tell us that Sun-like G-type stars can emit "superflares" with greater intensity than any other recorded solar flares (Maehara et al. 2012; Shibayama et al. 2013; Notsu et al. 2015a,b) with a frequency that is estimated as once every hundred years (Maehara et al. 2015). While it is still debated whether our Sun can cause superflares (e.g. Shibata et al. 2013; Aulanier et al. 2013), it is thought that solar flares and superflares are caused by the same mechanism (Karoff et al. 2016; Maehara et al. 2017; Namekata et al. 2017). Studies on cosmogenic radionucleocides show two cosmic-ray events in 774/775 and 993/994 by measuring carbon-14 in tree-rings (Miyake et al. 2012, 2013). They suggest a solar proton event as a candidate, as well as additional similar events in the Holocene (Miyake et al. 2017; Park et al. 2017; Wang et al. 2017; Fogtmann-Schulz et al. 2017). These contentions are also supported by ice core data (Thomas et al. 2013; Usoskin et al. 2013; Mekhaldi et al. 2015; Fogtmann-Schulz et al. 2017). They pose greater risk to modern civilisation than Carrington-class storms (Lingam & Loeb 2017). In either case, they are beyond the coverage of modern telescopic observations (e.g. Hoyt & Schatten 1998a,b; Hathaway 2010; Owens 2013; Clette et al. 2014; Vaquero et al. 2016; Svalgaard & Schatten 2016; Neidig & Cliver 1983a,b; Vennerstrom et al. 2016; Lefèvre et al. 2016; Riley et al. 2018).

However, historical documents let us trace further the history of solar flares and their associated magnetic storms by daily resolution with records of naked-eye sunspots and low-latitude auroras for at least millennia (Vaquero 2007a,b; Vaquero & Vázquez 2009; Willis & Stephenson 2001; Stephenson et al. 2004; Hayakawa et al. 2016c, 2017b). Historical documents of the Carrington storm in 1859 show that auroral displays were visible down to 23° MLAT (in a dipole geomagnetic model) in East Asia, and this result is consistent with the scale of magnetic disturbance reconstructed from occidental scientific literature (e.g. Kimball 1960; Tsurutani et al. 2003). Thus, historical documents in low MLAT areas are important to reconstruct and assess the scale of past magnetic storms. Historical documents can also be surveyed for cosmic-ray events in 774/775 and 993/994 (e.g. Usoskin et al. 2013; Stephenson 2015; Hayakawa et al. 2017a).

One of the Japanese chroniclers shows us an interesting description of another event in 1770 (Hayakawa et al. 2016b) that is also known to be the earliest aurora observation in both hemispheres (Willis et al. 1996; Nakazawa et al. 2004). This event even possibly rivaled the Carrington event. It has recently been reported that in 1770, a great sunspot at least twice as large as the one during the Carrington event possibly resulted in a series of great magnetic storms causing extremely bright low-latitude auroras visible down to 18.8° MLAT in association with precipitation of high-intensity low-energy electrons (HILEEs) to the upper atmosphere (Ebihara et al. 2017; Hayakawa et al. 2017e).

This historical evidence suggests that the Carrington event may not be so rare, with historical frequencies of about one event per century as suggested by theoretical and empirical discussions (e.g. Riley 2012; Curto et al. 2016; Riley & Love 2017; Riley et al. 2018).

Our question regards the existence of further evidence of extreme events within East Asian historical documents through the 18th-19th centuries, as this area is located at a considerably low MLAT (e.g. Butler 1992). One diarist recorded that the extreme auroral display in 1770 was not an unprecedented event at this low MLAT, recalling another auroral display at Kanazawa (see Hayakawa et al. 2017e, J091746).

This chronicler compared the "heaven's split" on September 17, 1770, with another one on February 6, 1728 (Kyoho 12, 12th month, 27th day). Although we could not find relevant records on this date, we found a great auroral display observed on February 15, 1730 (Kyoho 14, 12th month, 28th day; see Fig. 1(b)), supported with reports of simultaneous auroral observations in Japan (Osaki 1994; Nakazawa et al. 2004), China (Yau et al. 1995; Beijing Observatory 2008; Kawamura et al. 2016) and Europe (Fritz 1873; Angot 1896). These reports show that a significant space weather event was ongoing in 1730, and it is worth scaling this magnetic storm in comparison with contextual solar activity from the viewpoint of the solar-terrestrial relationship.

In this study, we conduct comprehensive surveys on contemporary historical documents from February 1730 in Japan, Korea, and China, and compare these accounts with discussions on contemporary solar activity. We focus our study on East Asia, where the observational sites had generally lower geomagnetic latitudes. We also compute the MLAT in every observational site in order to compare its scale with a Dst index for this event. Then, we examine the contemporary daily sunspot group counts to evaluate the solar activity around this date.

## 2. Method

In order to examine the great magnetic storm in 1730, we compare contemporary historical documents with luminous phenomena during nights in February 1730 in East Asian areas that were located at relatively low MLAT around 1730 (Butler 1992).

In China, we have not only official historical documents such as *Qīngshǐgǎo*, but also plentiful local treatises from various regions. These local treatises involve auroral records in their chapters of omens. In Korea, we have official historical documents and governmental diaries, *Joseon Wangjo Sillok* and *Seungcheonwon Ilgi*, for instance, written in the palace of the Joseon dynasty. In Japan, we consulted contemporary diaries written by people from various social classes. All the references to original historical documents are shown in Appendix 1.

We also examine sunspot observations recorded at about 1730 by contemporary astronomers such as Beyer, Weidler, Kraft, Hell, and Wasse, and we contextualise this historical magnetic storm with long-term solar activity reconstructed from their sunspot numbers (Hoyt & Schatten 1998a,b; Vaquero et al. 2016).

## 3. Overall result

Contemporary historical documents provide a total of 29 auroral records: 7 records in China and 22 records in Japan, but none in Korea. These records cluster around February 15, 1730, while European records report a series of auroral observations intermittently from February 3 to February 20 in the same year (e.g. Fritz 1873; Angot 1896). In the southern parts of Europe (e.g. Italy and the Iberian Peninsula), records also cluster around February 15, 1730. In Korea, we found no relevant records in Joseon Wangjo Sillok (v.24, ff.57b-58a) or the relevant folios of *Seungcheonwon Ilgi* (v.38, pp.620-637). We summarise the result in Table 1 and show their ID, observation date (year, month, date), colour, description, direction, time between start and end, their observation site (geographic latitude and longitude), and MLAT. The details of the observation sites and their MLAT are provided in the Discussion section.

The contemporary astronomers provide us with 37 sunspot observations in 1729 and 1730. These records are mainly from Beyer at Hamburg, Weidler at Wittenburg, Kraft at St. Petersburg, Hell at Peking, and Wasse at Northamptonshire (e.g. Vaquero et al. 2016). Within these sunspot observations, 24 records were previously available, 1 sunspot observation was not published previously, and 12 records have their values corrected with respect to a number of previously available records.

## 4. Discussion

### 4.1. Magnetic latitude of observation sites

It is known that the auroral oval boundary facing the equat corresponds to the scale of magnetic storms (Yokoyama et al. 1998; Tsurutani et al. 2003). Fortunately, we can estimate the lower limit of the scale based on locations of observation sites. In China, official historical documents are generally compiled in the contemporary capital city by the central government (e. Keimatsu 1976; Hayakawa et al. 2015) and explicitly declare the observation sites of auroral records, such as Fúshān (C1) and Barkul (C2). However, local Chinese treatises generally include the celestial and terrestrial omens observed in the very regions where these local treatises were compiled (C3-C6) or otherwise endorsed (C7). In Japan, the diaries show where the diarists themselves live, and hence the observation sites are considered to be where the diarists lived unless explicitly stated otherwise.

According to the historical magnetic field model GUF (Jackson et al. 2000), the magnetic north pole, where the dipole component of the geomagnetic field was located, was located at N 81.7°, E 307.1° in 1730. The observational sites of auroral records on February 15, 1730, range from Northern China to Northern Japan, between N35° and N44° in geographic latitude and between E93° and E141° in geographic longitude as shown in Fig. 2. We can compute the MLAT of these observational sites on the basis of the location of the contemporary magnetic north pole, as shown in Table 1. It should be noted that some Japanese records state that auroral displays were seen at the major cities Edo (N35° 42', E139° 45'), and Kamigata, i.e. Osaka (N34° 41', E135° 32'), and Kyoto (N34° 59', E135° 47'). We find only one record from Kyoto on this auroral display (J22), while we failed to find original records from Edo or Osaka so far. Hence we did not include Osaka and Edo as observation sites in this article.

According to the GUFM1 model, these observation sites are located at 25°-32° MLAT, except for that at Barkul (C2, 35.7° MLAT). Barkul is the westernmost observational site, approximately 19° westward in geographic longitude from the second westernmost observational sites (C3 and C4). The latitudinal and longitudinal difference between Barkul and the other sites probably indicates the latitudinal thickness and the longitudinal extent of the auroral oval. Purple vapour (C2) at Barkul reminds us of violet auroras at 427.8 nm ($N_2^+$ first negative band). One possibility for the purple vapour (C2) is a sunlit aurora, in which the high-altitude part of the upper atmosphere is directly illuminated by sunlight (Chamberlain 1961; Hunten 2003). Sometimes, high rays at wavelength 427.8 nm appear at higher altitude. Auroral emission at 630.0 [OI] is also enhanced relative to that at 557.7 [OI]. The sunlit aurora occurs at twilight, but C2 includes no information about the observation time. The sunlit aurora is very rarely described as a blue aurora. Tinsley et al. (1984) reported bluish low-latitude auroras "without red tint" seen at McDonald Observatory in Southwest Texas (N30° 41', W104° 01', 40° MLAT). They explained this phenomenon as the excitation of the $N_2^+$ first negative band by heavy neutrals or ions of H and O in the absence of sunlight. This type of aurora may be regarded as a 'ring current aurora', in which precipitating neutral H or $H^+$ excites the $N_2^+$ first negative band at low latitudes during large magnetic storms (Zhang et al. 2005).

### 4.2. Auroral extent and scale of the magnetic storm

The extent of auroral visibility does not necessarily mean that the equatorwardmost boundary of the auroral extent was seen in the zenith at either Kanazawa or other observational sites. Many historical documents with descriptions of auroral observations tell us that auroras were seen at directions between north-east and north-west, as shown in Table 1. Assuming that the upper part of the red aurora (400 km altitude) was seen at elevation angles of 45° at 27.4° MLAT (Kanazawa, J1, J9, J10, J11, J16, and J17), we can estimate that the equatorward boundary of the auroral oval would be located at 30.7° MLAT (33.5° invariant latitudes (ILAT). Here, we assumed that the aurora is distributed along the dipole magnetic field line. If the equatorward boundary of the auroral oval was located at 30.7° MLAT (33.5° ILAT), an observer at Kyoto (25.8° MLAT) could look at the aurora extending up to the elevation angle of 33°, and another observer at Hirosaki (31.5° MLAT) could see the same aurora near the zenith. This is consistent with the historical document at Hirosaki (31.5° MLAT) that the aurora was extended to the zenith (J13 and J19).

East Asian historical records provide us with a rough estimate of the elevation angle of auroral displays on February 15, 1730. According to the historical records, the auroral display "filled up the heaven" (C3-C5) or "filled up the ground" (C1, C7) in China and was described as a "great conflagration" (J2), "like fire between the east and north for several tens jou" (J16), and "half the sky in the north was as red as crimson" at Kaminoyama (29.1° MLAT, J20) in Japan. A report by Hirasawa Michiari (J13) excites our interest. He cites a report from Hirosaki and writes "In the Fief of Tsugaru, redness in the night was seen as southward in the Fief of Akita". Another record (J19: 31.5° MLAT) also reports that runners came from Tsugaru and thought that Akita or Ani was on fire, while they are southward from Tsugaru.

Assuming that the upper part of the red aurora (400 km altitude) was seen equatorward from Tsugaru (J19: 31.5° MLAT) and around the zenith of Kaminoyama (29.1° MLAT, J20), we consider the equatorward boundary of auroral oval would be located at 29.1° MLAT (32.1° invariant latitudes (ILAT)). The ILAT is the magnetic latitude at which a specific magnetic field line intersects the ground, and it is used to specify the magnetic field line of interest. For the dipole magnetic field, the ILAT is the same as MLAT on the ground. The MLAT is different from the ILAT above the ground. Here, we assumed that the aurora is distributed along the dipole magnetic field line. If the equatorward boundary of the auroral oval was located at 29.1° MLAT (32.1° ILAT), an observer at Kyoto (J22: 25.8° MLAT) could look at the aurora extending up to the elevation angle of 44.9°. If this is correct, the following scenarios will be plausible: An observer in Tsugaru did not see the aurora in the north direction, but did see the aurora extending from the southern horizon. This would mean that that the poleward edge of the red aurora was located at/or equatorward of Tsugaru. The exact location where the observer saw the aurora in the south is unclear. The upper limit of the poleward edge of the red aurora would be ~32° MLAT because the northernmost Tsugaru is located at ~32° MLAT). Following the description "half the sky in the north was as red as crimson" at Kaminoyama (29.1° MLAT, J20), the equatorward edge of the red aurora would be located at ~29° MLAT. If this is the case, the red aurora would be distributed from ~29° MLAT to ~32° MLAT, and the latitudinal width of the red aurora would be estimated to be ~3°. The latitudinal width of the red aurora is not unusual because the latitudinal width of HILEEs observed by the DMSP satellites during intense magnetic storms is about 1~3° in MLAT (Ebihara et al. 2017).

We calculated the 2D distribution of the auroral volume emission rates at 630.0 nm and 557.7 nm. The simulation scheme is the same as that performed by Ebihara et al. (2017), except that the flux of precipitating electrons was multiplied by 10, and the precipitation of the electrons was assumed to occur between 29° and 32° ILATs. The Spectrum B of precipitating electrons (Ebihara et al. 2017) was used to calculate the volume emission rate of the aurora, which corresponds to the HILEEs observed by the DMSP F8 satellite at 0130:08 UT on 14 March 1989 (Dst value of −589 nT). The result is shown in Fig. 3. The volume emission rate at 630.0 nm (red) dominates that at 557.7 nm (green). The observer located at 27.4° MLAT (Kanazawa) could see the aurora at elevation angles from ~40 to 110° from the northern horizon.

Another scenario is also plausible: Auroral display extended beyond the zenith of Tsugaru (31.5° MLAT; 34.2° ILAT) down to the zenith of Kaminoyama (29.1° MLAT, J20) in its maximum phase, but a northern part of the aurora was sometimes covered by cloud at Tsugaru. This scenario can explain the records that the aurora was seen only in the southern sky at Tsugaru, and that "golden light" was seen in the northern sky in the same place when the sky was clear enough (J3, J14, and J15). In both scenarios, the same auroras were able to be visible at Kyoto (25.8° MLAT, J22). Assuming the altitude of the aurora to be 400 km, we estimate the elevation angle to be 44.1°. Although the equatorward edge of the auroral oval is unclear, we tried scaling the Dst value of the magnetic storm based on the distance of the auroral extent from the equator. Applying the formula of Yokoyama et al. (1998) at 32.1° MLAT, we estimate the Dst value to be more extreme than −1200 nT. These values are more extreme than the most extreme magnetic storm in 1989 (−589 nT) based on formal Dst records since 1957, while it is much less extreme than the Carrington event in 1859 (e.g. Tsurutani et al. 2003; Nevanlinna 2008; Cliver & Dietrich 2013; Cid et al. 2014; Kumar et al. 2015; Lakhina & Tsurutani 2016). Of course, the estimation of the Dst value is not definitive because the ambiguity of the equatorward edge of the auroral oval and the deviation of the model of Yokoyama et al. (1998).

Contemporary reports from Western Europe provide supporting evidence for our estimation of the extension of the auroral oval. As shown by Kawamura et al. (2016), we also have contemporary auroral observations in Western Europe. Fig. 4(a) shows the auroral drawing reproduced from M. Caumette's report (t.1, pp.332-333 with its figure) at Marseille (N43° 18', E005° 28', 47.6° MLAT) on February 15, 1730. This auroral display was mainly whitish (*blanchâtre*) with some parts in bluish, reddish, or violet. This drawing clearly shows that the auroral display expanded from north to south. According to the GUFM1 model, we compute its magnetic inclination as 69.0° and magnetic declination as −13.4° at Marseilles in 1730 (Fig. 4(b)). We plot the contemporary magnetic zenith at Marseilles in Fig. 4(b). This figure shows that the concentric structure of the auroral display is roughly symmetric with respect to the magnetic meridian that extends from magnetic north and magnetic south through the magnetic zenith. Corona auroras, in which auroral rays seen near the magnetic zenith converge on the magnetic zenith (Chamberlain 1961), seem to be absent in the drawing. This probably implies that the greenish auroral display at the lower latitude border (100-120km) was dominant. The greenish band aurora probably appeared over Marseilles (47.6° MLAT) because the upward Region 1 field-aligned current can extend to at least ~47° MLAT, according to the DMSP satellite observation during the large magnetic storm of November 20, 2003 (Ebihara et al. 2005).

Weidlero and Rhostio (1730, pp. 4-5) showed us that the aurora borealis seen throughout Europe expanded even beyond the zenith and "inclined brilliantly from west to south" in Rome (N41° 53', E012° 29', 45.1° MLAT). The whole text can be translated as follows:

> The next year, i.e. 1730, cast a history of night lights to those who see, and have not given anything as attractive and attentive as such things to the curious writers. On the 15. day of February, the aurora borealis shone almost the whole night at Rome, Florence, Bern in Helvetia, Vienna in Austria, Bratislava, and at many other places. Especially, the heavenly region in Rome was inclined brilliantly from west to south and the spaces continued to be brilliant, although it is rare to be seen brilliance continuously. We read that also in Petersburg in Russia, the zenith, covered by a composed display, has been placed into rays dispersed by the spectrum.

Another report at Geneva (N46° 12', E006° 09'; 50.3° MLAT) by Cramer (1730) is especially intriguing within these occidental auroral observations. Cramer (1730 p.280) describes the auroral extent as follows:

> But what was chiefly to be considered, was a great Meridional Zone, pretty like a Rainbow in its Figure, but broader. It was terminated by two parallel Arches. The superior insisted with one Side upon the true Point of East, and with the other upon the Point of South-west, or West-south-west: Whence you see its Middle declined about 15 Deg. from South to East, and was diametrically opposed to the Middle of the Aurora Borealis. Its Altitude did vary a little, but never reached higher than the Head of Orion, which was 54 Deg. high, and never was seen lower than a little under Procyon, which is an Altitude of 45 or 46 Deg. The inferior Arch was exactly parallel to the superior, and the Breadth of the Zone varied from 14 or 15 Deg. to 18 or 20 Deg.

Cramer's report shows that the equatorward auroral boundary at Geneva on February 15, 1730, was "a little under Procyon" at 45° or 46° in elevation angle. In February, Procyon appears in the southern sky, and hence we consider that the auroral extent at Geneva surpassed the zenith to reach the southern sky at 45-46° elevation angle. Although it is not so frequent to see equatorward auroras surpassing the zenith in the mid-latitudes, it is possible to see auroras southward under extreme magnetic storms. During the Carrington magnetic storms, we have several reports for auroras observed "southward" in North America (e.g. Loomis 1860). We note that Cramer (1730) pointed out the rainbow-like structure of the auroral display that is sometimes the case with historical auroral records (e.g. Hayakawa et al. 2016a; Carrasco et al. 2017).

These facts suggest that the auroral oval on February 15, 1730, surpassed the zenith of Geneva (50.3° MLAT) and Rome (45.1° MLAT). These records are supported by Iberian auroral observations as well. On this date, auroras were seen in the Nisa region (45.9◦ MLAT) in Portugal and Seville (43.4° MLAT), and at Granada (42.9° MLAT) in Spain (Aragonès & Ordaz 2010).

### 4.3. Colour and brightness of auroras

While East Asian historical records do not explicitly state elevation angles of these auroral displays, their descriptions for auroral colour and brightness provide hints for this discussion. The descriptions (see, Table 1) show that the colours of this aurora varied considerably, while the auroral colour in East Asia during the Carrington magnetic storm was mostly red (Hayakawa et al. 2016b). Red auroras were dominant in this event as well (C1, C3-C5, C7, J4-J6, J12, J17-J18, and J20). In the same, we find purple auroras in Barkul (C2, 35.7° MLAT), gold-coloured auroras in Hirosaki (J3, J15, 31.5° MLAT), red auroras with white stripes in Hanamaki and Kubota (J7 and J19, 30.3° MLAT and 30.6° MLAT), and a five-coloured aurora in Nínghǎizhōu (C6-C7, 28.1° MLAT).

We mentioned the purple vapour (C2) above. Purple auroras were also reported in Geneva. Cramer (1730) states as follows: "The Colour of this Zone was Red, Scarlet, inclined to Purple, pretty lively and changeable by Intervals" (Cramer 1730, p.280). The case in Geneva is similar to one of the observations at Hirosaki (J2-J3). The diarists observed light like a "conflagration" and "golden light" in the northern portion of the sky. Lights like "conflagration" remind us of red auroras and OI emission at 630.0 nm that is frequently observed in low-latitude auroras. "Golden light" can be explained as a yellowish colour that has at least two reasons. One possibility is a mixture of bright red and green emissions. The other is selective scattering and attenuation of blue emission near the horizon (e.g. Hallinan et al. 1998). Red and white auroras observed at Hanamaki (J7) are typically seen in low-latitude auroras in great magnetic storms such as those seen in the Carrington event (Loomis 1860) or the great magnetic storms in 1770 (Nakazawa et al. 2004; Hayakawa et al. 2017e). "Ominous cloud with five colours" (C6-C7) should be a mixture of these various emissions. Great magnetic storms in 771/772 and 773 show auroras with blood-red, green, saffron, and black at Amida (MS Vat.Sir.162, f.150v, f.155v), and at 45° MLAT in contemporary time according to the Zūqnīn Chronicle (see Hayakawa et al. 2017b). Shiokawa et al. (1997) reported that green auroras down to 15° in MLAT on the polar side were observed from the ground station at Rikubetsu. Therefore, an auroral extension down to 43° MLAT is considered as the minimum condition to see auroras in purple, gold, or five colours, according to the study with modern instruments.

On the other hand, the descriptions of auroral brightness in these historical documents lead us to consider that these auroras were even more extreme than that of the Halloween event in 2003. Chinese historians describe these auroras as "filling up the ground" at Fúshān (C1), "filled up the heaven like a rising sun" in the Provinces of Shǎnxī and Shānxī (C3-C5), and "with five colours brilliantly and colourfully" at Nínghǎizhōu (C6, C7). Japanese diarists, except for J6, mostly compared these auroras with a conflagration. In particular, a Japanese chronicler at Hanamaki (J7) compared the auroral brightness on this date with sunset light. These brightness descriptions are totally different from appearances of normal low-latitude auroras with a dim red colour, which are can be difficult to capture with the naked eye. Cramer describes the aurora brightness at Geneva as "the Light was still, and clear enough to read a Character no bigger than that of this letter" (Cramer 1730, p. 280). In Nisa, it was reported: "There was such a clarity that people could see their own shade as they walked" (Aragonès & Ordaz 2010).

### 4.4. The Sun around February 1730

Unfortunately, the solar observations in 1730 suffer from a paucity of scientific data (Vaquero 2007a; Vaquero & Vázquez 2009). The monthly mean sunspot number in the Sunspot Index and Long-term Solar Observations (SILSO) provided by the World Data Center for the production, preservation, and dissemination of the international sunspot number at Brussels, is available only from 1749 (Clette et al. 2014), that is, the start of Staudach's sunspot drawings (Arlt 2008, 2009). On the other hand, yearly sunspot number estimates show that the nearest peak was in 1728, just before this event (Hoyt & Schatten 1998a,b; Clette et al. 2014; Carrasco et al. 2015a; Vaquero et al. 2016; Svalgaard & Schatten 2016). Hoyt & Schatten (1998a,b) and Vaquero et al. (2016) reported that Johann Beyer (1673-1751) made sunspot observations at Hamburg in early 1730. We highlight that Hermann Wahn pointed out that Beyer had a telescope in his observatory with a focal length of at least 3.6 m (Wahn 1731).

Wolf (1861b, p.82); Wolf (1861a, pp.115-116) revealed that the reports by Johann Beyer are found in *Niedersächsische neue Zeitungen von gelehrten* (hereafter, NNZG). The NNZG is a general journal with several sunspot drawings made by Beyer in the volumes corresponding to 1730 and 1731. The volume corresponding to 1731 contains five sunspot drawings dated to early 1730. Furthermore, we have found other sunspot drawings made by Beyer (not registered by Hoyt & Schatten 1998a,b) on December 14, 1729, where Beyer registered four sunspot groups (NNZG p. 836). In order to better understand the great magnetic storm that occurred in February 1730, we have carried out a revision of the sunspot group records between 1729 and 1731. Table 2 lists the observations available for that period, provides information about the observer, and indicates whether the sunspot record was published previously (N), corrected with respect to the group number shown in old sunspot group databases (C), or was previously available (O). Fig. 5 shows the sunspot group counts and date recorded by different astronomers during the period 1729–1731.

We highlight several novelties from this review. Weidler, in his observations, registered the number of single sunspots and not the group number (Carrasco et al. 2015a). However, from his annotations, the number of sunspot groups had been known for some time. Thus, we have obtained the group number observed by Weidler for two of his five observation days in 1729 (see Table 2). We have carried out a complete revision of sunspot records made by Beyer in 1729 and 1730 from the original documentary source (NNZG). We have recovered a new sunspot observation on December 14, 1729, which has not been published previously. Of the remaining 17 sunspot records, we have corrected the value for the number of sunspot groups in 10 cases, increasing the value in 9 cases (increments between 1 and 3 units) and decreasing, by one unit, the previous group number value in only one case on April 11, 1730. Moreover, the solar observations made by Beyer in February indicated in Hoyt & Schatten (1998a,b) and Vaquero et al. (2016) was actually performed in March. We also note that Beyer made 25 sunspot observations from April 14, 1730, to July 2, 1730. Beyer pointed out that the Sun had many sunspots between May and the end of June, and the number of sunspots greatly decreased after this period, while he only reported general information unfortunately (Wolf, 1873). On the other hand, we have revised the sunspot records by Adelburner in 1730. We consulted the original text written by Adelburner (Adelburner, 1735), and we can conclude that he himself did not observe sunspots but instead pointed out that sunspot observations were performed by Beyer (Adelburner, 1735, p. 80). Moreover, Adelburner indicates the number of single sunspots, not the group number. In fact, the two observation dates and the number of single sunspots reported by Adelburner are equal to what Beyer registered for these two observation days. For these reasons, Adelburner is not included in Table 2 or Fig. 5. Finally, Vaquero et al. (2007) demonstrated based on information compiled by Hell that the sunspot observation during the eclipse on July 15, 1730, was carried out by the Jesuits Kegler and Pereira, and not by Hallerstein. Moreover, the number of sunspot groups registered was three, not seven. Vaquero et al. (2007) also showed that the only solar observation available for 1731 compiled by Hell did not contain information about sunspots. In several cases, Hoyt & Schatten (1998a,b) assigned zero sunspots to solar records where information about the presence or absence of sunspots was not mentioned (Clette et al. 2014; Carrasco et al. 2015b). These changes in sunspot information compiled by Hell are corrected in the new revised collection of group numbers published by Vaquero et al. (2016).

However, we must admit that we cannot directly relate any of them to the great magnetic storm on February 15, 1730, as both of the nearest sunspot drawings are two weeks away from the date of the storm. Therefore, we consider that the sunspot in question was on the other side of the solar disc on January 27, 1730 (Fig. 6(a)) and March 1, 1730 (Fig. 6(b)) and hence escaped the eye of contemporary sunspot observers. In any case, Figure 5 shows a notable increase of the sunspot number in early 1730. This indicates that the general level of contemporary solar activity was very high, although the solar activity in the second part of the year was considerably lower (Carrasco et al. 2015a). It is known that auroras were intermittently observed in middle to Northern Europe between February 03 and February 20, 1730 (e.g. Fritz 1873; Angot 1896). These auroras may have been caused by clustering coronal mass ejections from the same active region, just like other clustering CMEs (e.g. Mannucci et al. 2005; Tsurutani et al. 2008; Hayakawa et al. 2017e).

## 5. Conclusion

We examined the historical records of a great auroral display that occurred on February 15, 1730. The auroral observations in East Asia were compared to contemporary South European reports. We can infer from these historical records that the auroras were visible at least down to 25.8° MLAT throughout East Asia. While the elevation angle for this auroral display cannot be computed exactly, the most equatorward boundary of the auroral display was confirmed at least beyond 45.1° MLAT and possibly down to 29.1° MLAT (32.1° ILAT) during the maximum phase of the geomagnetic storm. This auroral display was also very bright. We have considerable evidence from historical reports for non-red colour auroras, which suggests excitation of nitrogen atoms at high altitude as well.

The contemporary sunspot observations show that this event can be placed in the active phase of solar activity in the first half of 1730, after the maximum of the solar cycle occurred in 1728. An effort has been made to evaluate and improve our knowledge about the solar activity around 1730, correcting some erroneous values of the number of sunspot group observed, and adding other values in the database (Vaquero et al. 2016).

The study of major events of past space weather provides crucial data on these events, which occur rarely in each century (Riley 2012; Curto et al. 2016; Riley & Love 2017). Because of their rarity, these cases are of enormous interest to modellers of the magnetosphere, the heliosphere, and the disturbances that are propagated inside it. Thus, they could check their physical models by applying them to exceptional cases. Obtaining more data of these extreme space weather events will provide us further bases of physical understanding of low-latitude auroras under extreme space weather events (Cliver & Dietrich 2013). Moreover, the reconstructed sunspot number around 1730 will provide us important indications about the solar activity and development of active regions in the early 18th century, considering that we have limited information about the sunspot activity in the early 18th century before Staudach (Svalgaard & Schatten 2016; Vaquero et al. 2016). These actual data for long-term solar activity will have important consequences on the dynamo theory for the long-term evolution of solar activity (Vaquero et al. 2011; Augustson et al 2015; Hotta et al. 2016).

In conclusion, this magnetic storm can be considered one of the more interesting case studies of historical extreme event of the space climate. Based on our results, this event can be considered at least more intense than the great storm in 1989, but weaker than the famous Carrington event in 1859. Moreover, it was only 40 years before the extreme superstorms in 1770.

**Acknowledgements**.
We thank Mr. K. Fukuda for helpful comments on the interpretation of the occidental texts, K. Mase for her helpful advices on Japanese historical records, SILSO for providing sunspot data, T. Takeda for his helpful advice on Chinese historical documents, and Y. Oku for his help in accessing French records. We thank the Kanazawa Tamagawa Library, the Hirosaki City Library, the Archives of the Institute of Research for Humanity of Kyoto University, and other archives for letting us survey their historical documents. We gratefully acknowledge the support of Kyoto University's Supporting Program for the Interaction-based Initiative Team Studies "Integrated study on human in space" (PI: H. Isobe), the Interdisciplinary Research Idea contest 2014 held by the Center for the Promotion of Interdisciplinary Education and Research, the "UCHUGAKU" project of the Unit of Synergetic Studies for Space, the Exploratory and Mission Research Projects of the Research Institute for Sustainable Humanosphere (PI: H. Isobe) and SPIRITS 2017 (PI: Y. Kano) of Kyoto University, and the Center for the Promotion of Integrated Sciences (CPIS) of SOKENDAI. This work was also encouraged by Grant-in-Aids from the Ministry of Education, Culture, Sports, Science and Technology of Japan, Grant Num- ber JP15H05816 (PI: S. Yoden), JP15H03732 (PI: Y. Ebihara), JP16H03955 (PI: K. Shibata), and JP15H05815 (PI: Y. Miyoshi), and a Grant-in-Aid for JSPS

Research Fellow JP17J06954 (PI: H. Hayakawa). This research was also partially supported by the Economy and Infrastructure Counselling of the Junta of Extremadura through project IB16127 and grant GR15137 (co-financed by the European Regional Development Fund) and by the Ministerio de Economía y Competitividad of the Spanish Government (AYA2014-57556-P and CGL2017-87917-P).

**Appendix A: Supplements**

In these supplements, we provide the list of historical sources. They are shown in the original language of these historical sources, in order to keep them traceable. Historical sources in Appendix A.1.1 are in Chinese (C) and Japanese or Sino-Japanese (J), and those in Appendix A.1.2 are in European languages. For texts in wood/lead prints, we provide the romanised title, original title, volume number (v.), and folio numbers (f.). For texts in manuscripts, we provide the romanised title, original title, volume number, folio number, location, and shelf mark. For texts from published critical editions, we provide the romanised title, original title, page number, name of publication, and year of publication.

**Appendix A.1: References of historical sources**

**Appendix A.1.1: Historical sources for East Asian auroral observations**

- C1: Qīngshǐgǎo, 清史稿, v.41, p.1572
- C2: Qīng Shìzōng Shílù, 清世宗実録, v.103, f.22b
- C3: Qīng Yōngzhèng Shānxī Shuòpíngfǔzhì, 清雍正 山西朔平府志, v.11, f.17b
- C4: Qīng Yōngzhèng Shānxī Shuòzhōuzhì, 清雍正 山西朔州志, v.2, f.45a
- C5: Qīng Qiánlóng Shǎnxī Fǔgǔxiànzhì, 清乾隆 陝西府谷縣志, v.4, f.1b
- C6: Qīng Tóngzhì Shāndōng Nínghǎizhōuzhì, 清同治 山東寧海州志, v.1, f.24a
- C7: Qīng Guāngxù Zēngxiū Dēngzhōufǔzhì, 清光緒 増修登州府志, v.23, f.11b
- J1: Ogura Nikki, 小倉日記, p.326; 日本海地域史料叢書第一冊 一丸諸事記 附小倉日記抄, 1984
- J2: Eiroku Nikki, 永禄日記, p.160; 青森県文化財保護協会, みちのく叢書, v.1, 1983
- J3: Hirayama Nikki, 平山日記, p.223; 青森県文化財保護協会, みちのく叢書, v.17, 1983
- J4: Kakan Shousetsu, 可観小説, v.26, p.479; 金沢文化協会, 可観小説後編, 1936

- J5: Shin'nenji Kibocho Sho, 真念寺鬼簿帳抄 『柳田村史』p.1388, 1975.
- J6: Hanamaki Nenkei, 花巻年契 『花巻市史（年表史料）』p.12, 1970.
- J7: Kain, 花印 『花巻市史（年表史料）』p.120, 1970.
- J8: Kain, 花印 『花巻市史（年表史料）』p.120, 1970.
- J9: Arisawa Takesada Nenpu, 有沢武貞年譜, v.2, f.55b-56a; 金沢市立玉川図書館 近世史料館 MS 特 16.38-2-6
- J10: Tooda Nikki, 遠田日記, v.5, ff.4b-5a; 金沢市立玉川図書館 近世史料館 MS 特 16.42-5-9
- J11: Gokokukou Nenpu, 護国公年譜, v.3, f.22b; 金沢市立玉川図書館 近世史料館 MS 特 16.11-3-86
- J12: Satsuyou Sado Nendaiki, 撮要佐渡年代記, p.105; 佐渡叢書, v.4, 1973.
- J13: Hirasawa Michiari Nikki, 平沢通有日記, v.2, p.122; 秋田市歴史叢書, v.2, 2008.
- J14: Umeda-mura Hikoroku Kaki, 梅田村彦六家記, p.134: 奥瀬清簡『本藩旧記』v.1, 1980
- J15: Honpan Min Jitsuroku, 本藩明實録, p.294; みちのく双書, v.45, 2002.
- J16: Honpan Rekifu, 本藩暦譜, p.69; 金沢市史, v.3, 1999.
- J17: Sanshu Chirishi, 三州地理志, v.1, p.13; 三州地理志, 1931.
- J18: Kiji Besshu, 記事別集, p.40, 越佐叢書, v.12, 1977.
- J19: Kami-Sakana-cho Kiroku, 上肴町記録, p.69; 秋田市史, v.9 近世史料編上巻, 1997.
- J20: Kaminoyama Sanke Kenbun Nikki, 上山三家見聞日記, p.113, A433; 資料編第十七集, 1976.
- J21: Zoku Zentoku Zakki, 続漸得雑記, v.15, ff.88b-89b; 金沢市立玉川図書館 近世史料館 MS 特 16.05-15-006
- J22: Myouhouin Nikki, 妙法院日次記, v.6, p.418; 妙法院日次記, 1989.

**Appendix A.1.2: Historical sources for European auroral observations**

- Caumette, M. 1766, Description d'un phénomène observé à Marseille, le 15 février 1730, *Histoire de la Société royale des sciences établie à Montpellier*, t.1, pp.332-333
- Cramer, M. 1730, An Aurora Borealis attended with unusual Appearances, in a Letter from the Learned Mr. G. Cramer, Prof. Math. Genev. to James Jurin, M.D. and F.R.S., *Philosophical Transactions*, 36, 413, 279-282.
- Weidlero, J. F., Rhostio, C. S. 1731, *De Meteoro Lucido Singulari a. MDCCXXX. M. Octobri Conspecto Dissertatio qua Observationes Madritensis et Vitembergensis inter se Comparantur.*

*Gerdesii*, Vitembergae.

**Appendix A.1.3: Historical sources for sunspot observations**

- Adelburner, 1735, *Commercium Litterarium ad Astronomiae Incrementum*. Norimbergae.
- Anonymous 1731, *Niedersächsische neue Zeitungen von gelehrten*, Göttingen.
- Hell, M. 1768, *Observationes Astronomicae ab anno 1717 ad Annum 1752 Pekini Sinarum Factae*, Vindobonae.
- Wahn, H. 1731, *Staatskalender*, Hamburg.
- Weidler, J.F. 1729, *Observationes Meteorologicae Atque Astronomicae*, Vitembergae.

**Appendix A.2: Selected records with transcriptions and translations**

We provide selected translations of original historical sources. We provide a translation of texts with information about an equatorward extension of the auroral display. The exact references of these documents are available in Appendix A.1.

[Note: Please see the publisher version for this section]

**Figures**

**Figure 1**

[Note: Please consult the publisher version for Figures 1a and 1b]

Fig. 1. (a) Seirinki (MS 16.28-11-11) in Kanazawa Tamagawa Library, with a description of "heaven's split" in September 17, 1770 (Courtesy: Kanazawa Tamagawa Library). (b) Japanese auroral record on February 15, 1730: Arisawa Takesada Nenpu, v.2, f.55b-56a; MS toku 16.38-2-6 in Kanazawa Tamagawa Library (Courtesy: Kanazawa Tamagawa Library). This record is registered as J9 in our paper (see the Supplementary Information and Table 1).

**Figure 2**

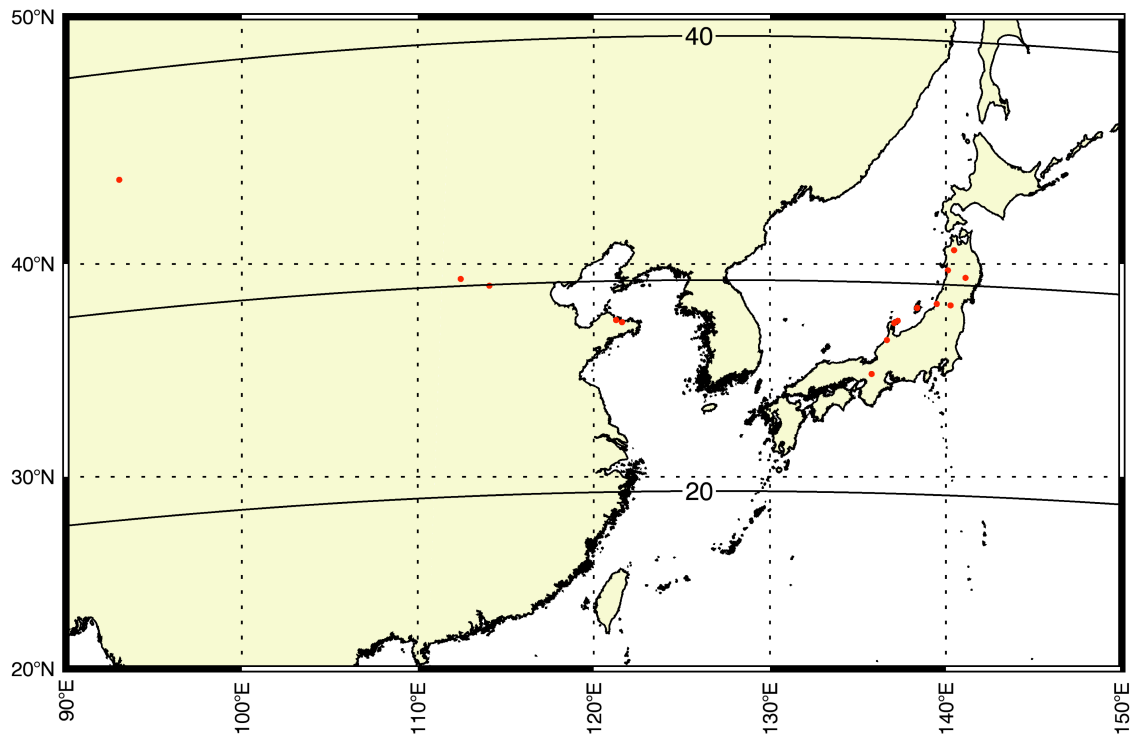

Fig. 2. East Asian observational sites of the great auroral display on February 15, 1730. The thick lines indicate the magnetic latitudes at 20°, 30°, and 40°.

**Figure 3**

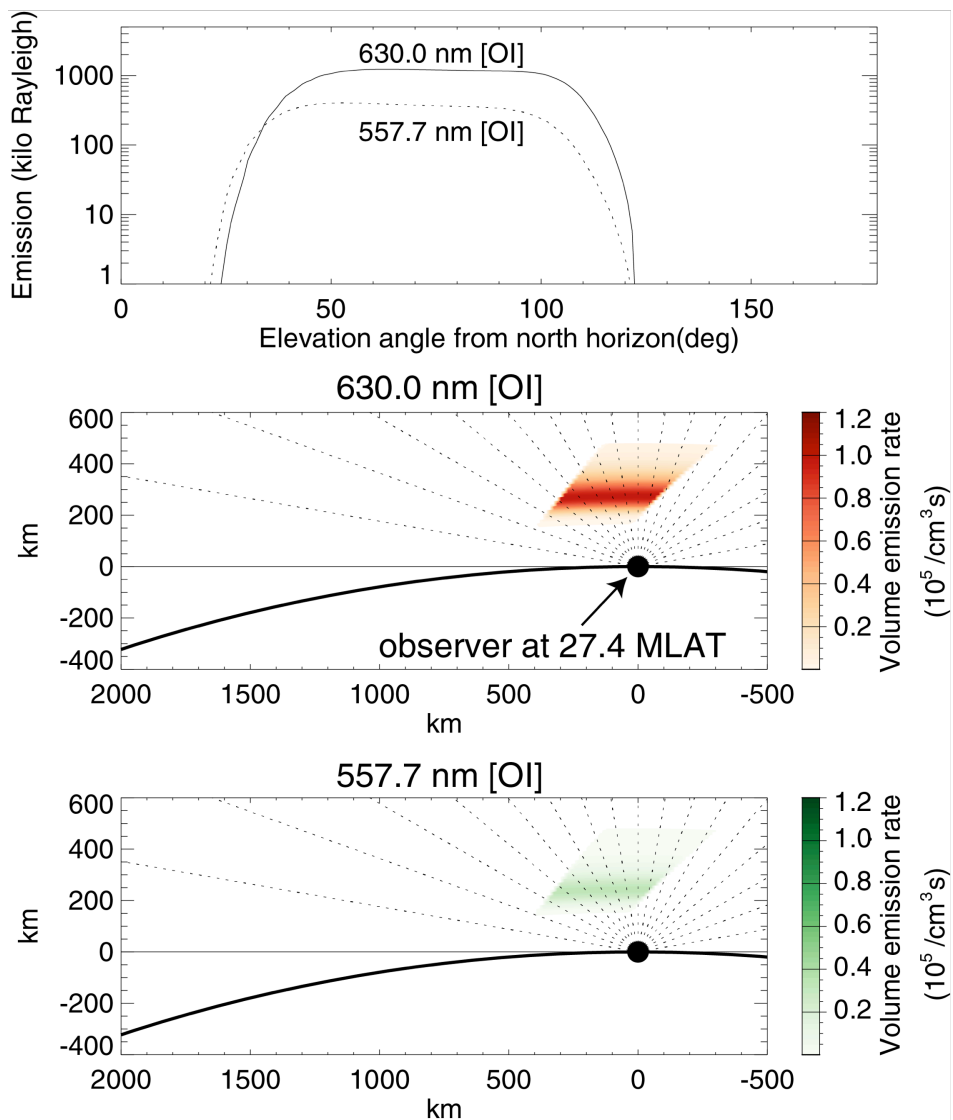

Fig. 3. Calculated auroral emission for the scenario in which the electron precipitation is assumed to occur between 29° and 32° ILATs. (top) Column emission rates at 630.0 nm (solid) and 557.7 nm (dotted) from the point of an observer at 27.4° MLAT (black circle), (middle) the volume emission rate at 630.0 nm in the meridional plane, and (bottom) the volume emission rate at 557.7 nm. The electron precipitation is assumed to occur between 29° and 32° ILATs. The solid curve indicates the surface of the Earth. Magnetic north is to the left. The red aurora dominates the green one. The simulation result may be consistent with the description "half the sky in the north was as red as crimson" (J20).

**Figure 4**

[Note: Please consult the publisher version for Figures 4]

Fig. 4. (a) Auroral drawing by M. Caumette upon observation at Marseilles. (b) The magnetic zenith of contemporary Marseilles. (c) Integrated imagery of the auroral drawing (a) and magnetic zenith (b) at Marseilles on February 15, 1730.

**Figure 5**

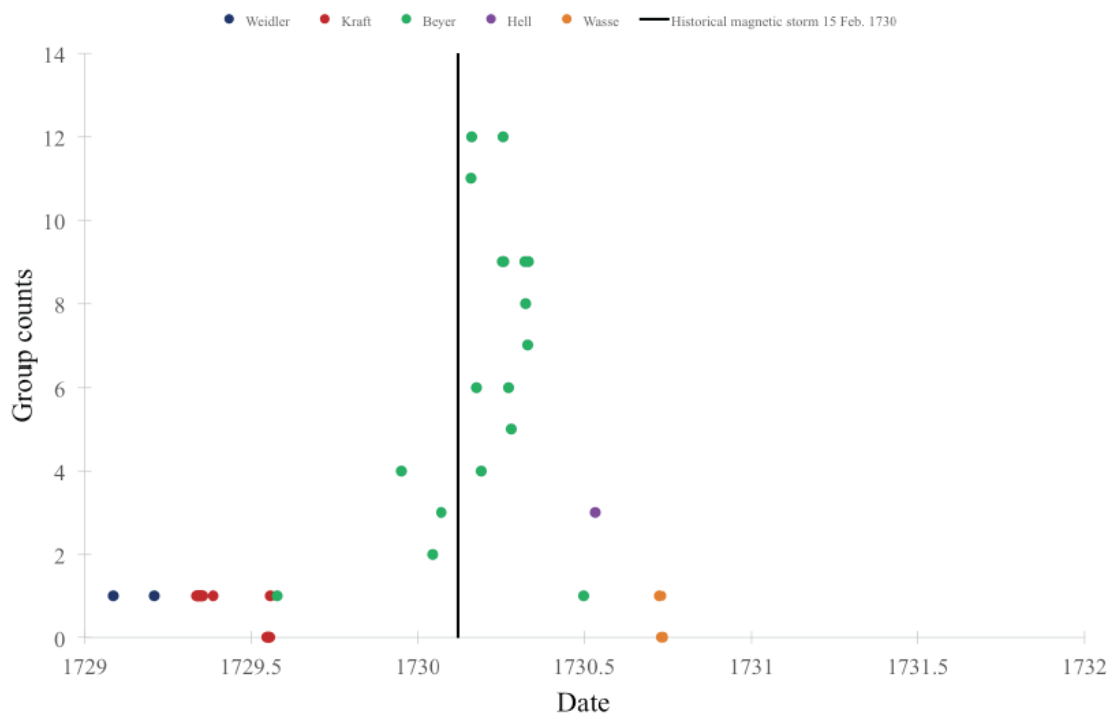

Fig. 5. Sunspot group counts from 1729 to 1731, based on Table 2. Blue, red, green, purple, and orange represent the solar observations made by Weidler, Kraft, Beyer, Hell, and Wasse, respectively. The vertical black line depicts the date when a historical storm studied in this work occurred (February 15, 1730).

**Figure 6**

[Note: Please consult the publisher version for Figure 6]

Fig. 6. Contemporary sunspot drawings by Beyer on (a) January 27, (b) 1730 and March 1730.

**Tables**

Table 1. Records of auroral observations on February 15, 1730

| ID | Year | Month | Date | Colour | Direction | Start | End | G. Lat. | G. Long. | MLAT |
|---|---|---|---|---|---|---|---|---|---|---|
| C1 | 1730 | 2 | 15 | R | | | | N37° 29' | E121° 16' | 28.2 |
| C2 | 1730 | 2 | 15 | Pu | en | | | N43° 36' | E093° 02' | 35.7 |
| C3 | 1730 | 2 | 15 | R | n | 22:00 | 26:00 | N39° 20' | E112° 26' | 30.3 |
| C4 | 1730 | 2 | 15 | R | n | 22:00 | 26:00 | N39° 20' | E112° 26' | 30.3 |
| C5 | 1730 | 2 | 15 | R | n | 22:00 | 26:00 | N39° 02' | E114° 04' | 30 |
| C6 | 1730 | 2 | 15 | FC | | 22:00 | | N37° 23' | E121° 36' | 28.1 |
| C7 | 1730 | 2 | 15 | R | | | | N37° 29' | E121° 16' | 28.2 |
| C7 | 1730 | 2 | 15 | FC | | 22:00 | | N37° 23' | E121° 36' | 28.1 |
| J1 | 1730 | 2 | 15 | | e-n | | | N36° 34' | E136° 39' | 27.4 |
| J2 | 1730 | 2 | 15 | Gl | wn | | | N40° 36' | E140° 28' | 31.5 |
| J3 | 1730 | 2 | 15 | Gl | wn | | | N40° 36' | E140° 28' | 31.5 |
| J4 | 1730 | 2 | 15 | R | wn-en | 20:00 | 28:00 | N36° 34' | E136° 39' | 27.4 |
| J5 | 1730 | 2 | 15 | R | | | | N37° 21' | E137° 04' | 28.2 |
| J6 | 1730 | 2 | 15 | R | | 22:00 | 26:00 | N39° 23' | E141° 07' | 30.3 |
| J7 | 1730 | 2 | 15 | R-W | wn-en | 22:00 | 26:00 | N39° 23' | E141° 07' | 30.3 |
| J8 | 1730 | 2 | 15 | | | daytime | night | N40° 36' | E140° 28' | 31.5 |
| J9 | 1730 | 2 | 15 | R | en-w | | 28:00 | N36° 34' | E136° 39' | 27.4 |
| J10 | 1730 | 2 | 15 | | e-n | | | N36° 34' | E136° 39' | 27.4 |
| J11 | 1730 | 2 | 15 | | n | 22:00 | | N36° 34' | E136° 39' | 27.4 |
| J12 | 1730 | 2 | 15 | R | | 22:00 | 28:00 | N38° 02' | E138° 22' | 28.9 |
| J13 | 1730 | 2 | 15 | R-W | e-w | 22:00 | | N39° 43' | E140° 07' | 30.6 |
| J13 | 1730 | 2 | 15 | | s | | | N40° 36' | E140° 28' | 31.5 |
| J14 | 1730 | 2 | 15 | Gl | wn | | | N40° 36' | E140° 28' | 31.5 |
| J15 | 1730 | 2 | 15 | Gl | wn | | | N40° 36' | E140° 28' | 31.5 |
| J16 | 1730 | 2 | 15 | | en | | | N36° 34' | E136° 39' | 27.4 |
| J17 | 1730 | 2 | 15 | R | wn-en | 20:00 | 28:00 | N36° 34' | E136° 39' | 27.4 |
| J18 | 1730 | 2 | 15 | | n | 20:00 | dawn | N38° 13' | E139° 29' | 29.1 |

| J19 | 1730 | 2 | 15 |     | wn    | 20:00 |       | N39° 43' | E140° 07' | 30.6 |
| J19 | 1730 | 2 | 15 |     | s     |       |       | N40° 36' | E140° 28' | 31.5 |
| J20 | 1730 | 2 | 15 | R   | wn-n  |       |       | N38° 09' | E140° 16' | 29.1 |
| J21 | 1730 | 2 | 15 | R-W | e-n   | 20:00 | 29:00 | N37° 27' | E137° 16' | 28.3 |
| J22 | 1730 | 2 | 15 |     | n-nne | 26:00 | dawn  | N34° 59' | E135° 47' | 25.8 |

Notes. The colour is described as R (red), W (white), Pu (purple), Gl (golden), or FC (five colours) in these historical documents. The direction is given by the eight points of the compass. Their observational time is given from 06:00 to 30:00 (06:00 on the following day) and time beyond the midnight is given as local time +24 on the same day, in order to categorise the observations that lasted beyond midnight in the same record. Their source documents are shown in the Appendix.

Table 2. Sunspot group observations available for the period 1729–1731.

| Observer | Date | Groups | Type | Observer | Date | Groups | Type |
| --- | --- | --- | --- | --- | --- | --- | --- |
| Weidler | 1729-2-1 | 1 | O | Weidler | 1729-3-18 | 1 | O |
| Kraft | 1729-5-3 | 1 | O | Kraft | 1729-5-4 | 1 | O |
| Kraft | 1729-5-5 | 1 | O | Kraft | 1729-5-6 | 1 | O |
| Kraft | 1729-5-7 | 1 | O | Kraft | 1729-5-8 | 1 | O |
| Kraft | 1729-5-9 | 1 | O | Kraft | 1729-5-10 | 1 | O |
| Kraft | 1729-5-22 | 1 | O | Kraft | 1729-7-20 | 0 | O |
| Kraft | 1729-7-21 | 0 | O | Kraft | 1729-7-22 | 0 | O |
| Kraft | 1729-7-23 | 0 | O | Kraft | 1729-7-24 | 1 | O |
| Beyer | 1729-7-31 | 1 | O | Beyer | 1729-12-14 | 4 | N |
| Beyer | 1730-1-17 | 2 | O | Beyer | 1730-1-27 | 3 | O |
| Beyer | 1730-3-1 | 11 | C | Beyer | 1730-3-2 | 12 | C |
| Beyer | 1730-3-7 | 6 | C | Beyer | 1730-3-12 | 4 | C |
| Beyer | 1730-4-4 | 9 | C | Beyer | 1730-4-5 | 12 | C |
| Beyer | 1730-4-6 | 9 | O | Beyer | 1730-4-11 | 6 | C |
| Beyer | 1730-4-14 | 5 | O | Beyer | 1730-4-29 | 9 | C |
| Beyer | 1730-4-30 | 8 | C | Beyer | 1730-5-2 | 7 | C |
| Beyer | 1730-5-3 | 9 | C | Beyer | 1730-7-2 | 1 | O |
| Hell | 1730-7-15 | 3 | C | Wasse | 1730-9-23 | 1 | O |
| Wasse | 1730-9-25 | 0 | O | Hell | 1731-12-29 | - | C |

Notes. The three first columns show the observer, date, and number of sunspot groups registered. The last column indicates if a sunspot record was previously available (O), if a sunspot observation was not published previously (N), or if a value was corrected with respect to group numbers of previously available records (C). In total, 37 observations are available during this time span in addition to Hell's record.